\documentclass[]{spie}  %>>> use for US letter paper
\usepackage{amsmath,amsfonts,amssymb}
\usepackage{graphicx}
\usepackage[colorlinks=true, allcolors=blue]{hyperref}

\title{ An Ultraviolet imager to study bright UV sources }

\author[1]{Joice Mathew}
\author[1]{Ajin Prakash}
\author[1]{Mayuresh Sarpotdar}
\author[1]{A.G. Sreejith}
\author[1]{Margarita Safonova}
\author[1]{Jayant Murthy}

\affil[1]{Indian Institute of Astrophysics, Bangalore, India}

\authorinfo{Further author information: (Send correspondence to Joice Mathew)\\E-mail: joice@iiap.res.in}
\begin{document} 
\maketitle

\begin{abstract}
We have designed and developed a compact ultraviolet imaging payload to fly on a range of possible platforms such as high altitude balloon experiments, cubesats, space missions, etc. The primary science goals are to study the bright UV sources (mag $< 10$) and also to look for transients in the Near UV (200 - 300 nm) domain. Our first choice is to place this instrument on a spacecraft going to the Moon as part of the Indian entry into Google lunar X-Prize competition. The major constraints for the instrument are, it should be lightweight ($< 2 Kg$), compact (length $< 50 cm$) and cost effective. The instrument is an 80 mm diameter Cassegrain telescope with a field of view of around half a degree designated for UV imaging. In this paper we will discuss about the various science cases that can be performed by having observations with the instrument on different platforms. We will also describe the design, development and the current state of implementation of the instrument. This includes opto-mechanical and electrical design of the instrument. We have adopted an all spherical optical design which would make the system less complex to realize and a cost effective solution compared to other telescope configuration. The structural design has been chosen in such a way that it will ensure that the instrument could withstand all the launch load vibrations. An FPGA based electronics board is used for the data acquisition, processing and CCD control. We will also briefly discuss about the hardware implementation of the detector interface and algorithms for the detector readout and data processing.  
\end{abstract}


\section{Introduction}
\label{sect:intro} 

The UV domain still remains largely unexplored: interesting and unique science may come through observations in this wavelength band. Because of absorption in the Earth's atmosphere, UV observations are possible only from space or from near space. There are now several initiatives in place for launches into near space and beyond, such as high-altitude balloon experiments, sounding rockets, LEO satellites, space missions, etc.\cite{B1}. Our goal is to develop a UV payload ready to fly on a range of such possible platforms.

%Most of the previous and existing UV space missions do not observe bright objects (NUV mag $< 10$) because of the detector safety constraints (Cao et al. 2011). 
Most of the previous and existing UV space missions have severe brightness limits (NUV mag $< 10$) due to the detectors 
safety constraints \cite{lut} and, therefore,
bright objects are excluded from observations. Our main scientific goal is to observe bright objects in the NUV (200 - 300 nm), which include hot massive stars, bright transients, solar system bodies, etc. \cite{luci}. Small space telescopes are able to provide significant UV science and can be realized in a cost effective way \cite{B2}. The main design considerations for the instrument are the capability to observe the bright sources in the NUV band, a wide field of view (FOV), compactness, light weight, cost-effective development and space qualification.

% \subsection{Use of This Document}

\section{Science Objectives}

The main scientific objectives of the instrument are studies of bright stars, variability in UV domain and transient observations.  The instrument can be used to acquire time-series of photometry in the (200-300 nm) range to study the variability  and environment of bright stars. Based on different flying platforms different science cases can be derived and this is described in the following sections.
\section{Flight Opportunities}

\subsection{High Altitude Balloon Experiments}
We are planning to fly this telescope onboard the high-altitude balloon to perform observations in the NUV domain from an altitude of around 30 km, mainly to look at bright UV sources, and also to look at the planetary bodies. Our team has developed an attitude sensor and stabilization platform for the  balloon-born astronomical payloads (Refs.~\citenum{5},~\citenum{6}).
A pointing system which carries out a coarse pointing and fine pointing is under development\cite{pointing}. And this will allow us to observe comets and other transients, such as supernovae in bands not observable by any other mission.The instrument will be mounted on this pointing platform for observations. By using this high altitude platform the instrument can be used to observe bright UV sources, solar system bodies, comets etc. Surprisingly for such a prominent object in the sky, there are only a handful of UV observations of the Moon\cite{moon}, primarily because the Moon is so bright that most spacecraft have an avoidance zone around it. We propose to observe both spectroscopically and with an imager to track changes in the albedo at different phases, where the phase angle of the Solar illumination changes. The Moon has garnered increased interest because of the renewed interest from space agencies and it is important to understand the nature of the Lunar surface.

\subsection{Nano Satellites}
Our instrument can be send as a scientific payload in nano satellites too since its instrument dimension is comparatively small.  Most of the bright stars are hot and they emit most of their radiation in UV. The photometric study of these bright stars helps to understand the different kinds of variability.The UV imager can be incorporated as a nano satellite payload in the LEO orbit and can perform studies of variability of bright stars, with focus on hot stars of spectral types O and B. 

\subsection{Lunar Platform}
Currently, we are in a collaboration with the Team Indus\footnote{The official website is http://blog.teamindus.in}, an Indian contestant for the Google Lunar X~PRIZE competition\footnote{The Google Lunar XPRIZE is the international competition with \$40 million incentive sponsored by Google and operated by the XPRIZE Foundation; official website {\tt http://www.googlelunarxprize.org/}.},  to send this UV telescope -- Lunar Ultraviolet Cosmic Imager (LUCI) -- to the Moon. LUCI will be mounted on the lunar lander and will perform a survey of the sky primarily looking for bright UV transients \cite{luci}. Observations from the Moon provide a unique opportunity to observe the sky from a stable platform far above the Earth atmosphere.

LUCI will be stored in a storage bay on the Lander horizontally. The storage bay will protect LUCI from dust
during non-operational period. LUCI will be deployed towards zenith from its horizontal position during its
operational period. Since LUCI is planned to be mostly contained within the lander, we will need only
minimal thermal blankets, since the platform will take care of most of the insulation, and the situation is similar
for the radiation shielding. The aperture will be well baffled to avoid reflections and scattering from the outer
surfaces of the lander. In order to ensure the survival over the lunar night, the telescope will be lowered back
into its storage bay and the detector will be switched off. The platform will have a minimal battery power to
provide the heater.
%\section{Instrument design consideration}

\section{Instrument Overview}

The instrument is a spherical catadioptric telescope with 200 -- 300 nm passband. It has a rectangular field of view of $0.46^{\circ}\times 0.34^{\circ}$  and the aperture diameter of 80 mm. The specifications of the instrument are given in Table.~\ref{table:instrument details}.

% \begin{figure}[h]
% \begin{center}
% \includegraphics[scale=0.7]{luci_system.png} 
% \end{center}
% \caption{LUCI block diagram}
% \label{fig:LUCI block diagram}
% \vskip 0.5in
% \end{figure}

\begin{table}[h]
\caption{Instrument Details} 
%\label{tab:fonts}
\begin{center} 
\begin{tabular}{l ll} 
\hline  
\rule[-1ex]{0pt}{3.5ex} Instrument & UV Imager \\
\rule[-1ex]{0pt}{3.5ex} Telescope & Spherical catadioptric  \\ 
\rule[-1ex]{0pt}{3.5ex} Field of View & $0.46^{\circ}\times 0.34^{\circ}$ \\
\rule[-1ex]{0pt}{3.5ex} Geometrical Collecting Area   &  $\sim 30 $  cm$^2$   \\ 
\rule[-1ex]{0pt}{3.5ex} Focal length & 800.69 mm  \\
\rule[-1ex]{0pt}{3.5ex} Detector & UV-sensitive CCD \\ 
\rule[-1ex]{0pt}{3.5ex} Sensor format & $1360 {\rm (H)}\times1024 {\rm (V)}$ pixels ($4.65\times 4.65\,\mu$m)   \\
\rule[-1ex]{0pt}{3.5ex} Pixel Scale & $1.2^{\prime\prime}$/pixel \\
\rule[-1ex]{0pt}{3.5ex} Resolution & $\sim 5^{\prime\prime} $\\
\rule[-1ex]{0pt}{3.5ex} Band of operation & $200-300$ nm\\
\rule[-1ex]{0pt}{3.5ex} Weight & $< 2$ kg \\
\rule[-1ex]{0pt}{3.5ex} Dimension & $450 \times 150$ mm (L $\times$ D)\\
\rule[-1ex]{0pt}{3.5ex} Power & $ < 10$  W \\
\hline
\end{tabular}
\label{table:instrument details}
\end{center}
\end{table}

All optical elements of the telescope are spherical. This includes primary and secondary mirror coated with AlMgF$_2$ and one Fused silica corrector lens . We have designed the telescope structure in such a way that it should have as minimal mass as possible and still be able to handle various types of launch loads and landing vibrations. The material we have chosen primarily is Aluminum, which is inexpensive, easily available, easy to handle and has a good strength. For the primary and secondary mirror mount, we have chosen Invar because of its very low thermal expansion properties, and titanium as material for fasteners.

A tubular structure with ribs has been used to ensure better stiffness and strength. Heating strips will be provided along the ribs to maintain the working temperature of the electronics and the whole instrument will be covered with Multi-Layered Insulation (MLI) to protect the telescope and the electronic parts from solar radiation and heat.

The detector is a UV-sensitive CCD (Charge Coupled Device)  from ARTRAY with the response between 200 -- 900 nm, therefore we have used a filter to restrict the bandpass to 200 -- 300 nm. An FPGA-based electronics module will be used to acquire the data from the detector and for the subsequent onboard processing and compression.

\section{Optical design}

The recent trend in building wide-field telescope is to use the RC-based telescope design with many corrector lenses. Our main design goal is to come up with a cost-effective alternative design which would consist of only spherical surfaces and a minimum number of optical elements. The advantages of telescope configurations  employing only spherical surfaces are the ease of manufacturing and alignment, and a low cost. The main feature of the optical system making it a cost-effective solution is that it consists of spherical primary and secondary mirrors and one meniscus double-pass lens, which makes the system less complex to realize. Since both mirrors are spherical, additional lens elements are required to further correct the aberrations. We have used only one single meniscus lens as a  corrector in the optical system, for simplicity.

\begin{figure}[h]
\begin{center}
\includegraphics[scale=0.80]{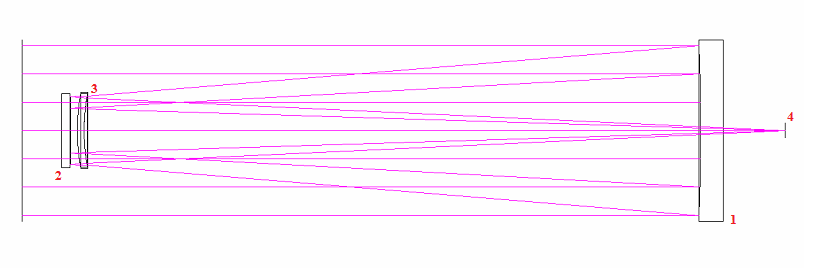} 
\end{center}
\caption{Optical layout: 1. primary mirror 2. secondary mirror 3. lens 4. focal plane}
\label{fig:optical layout}
\end{figure}

The optical system consists of two spherical mirrors and one spherical lens near the secondary mirror. The optical layout is as shown in (Fig.~\ref{fig:optical layout}) Since both mirrors are spherical, a lens element is used to correct for the spherical aberration and other off-axis aberrations. The optical design is carried out using the optical design software Zemax\footnote{Zemax is the software for the design and optimization of optical systems. The official website is http://www.zemax.com/}.
%The radius of curvatures of the optical elements, the spacing between the elements are the main variables we have used for the optimization of the design.
The main variables we have used for the optimization of the design are the curvature radii of the optical elements and the spacing between the elements.

The optical system has a focal length of 800.69 mm with F-number 10.67. The primary mirror is 80 mm diameter and has a radius of curvature of 914.30 mm. The primary mirror will be mounted on a blade cell mount made from Invar. The secondary mirror has a diameter of 33.15 mm and 865.53 mm radius of curvature. Both mirrors will be made from Zerodur, with an Al reflective surface and a protective MgF$_2$ coating to preserve the reflectivity in the near-UV. One corrector lens is placed just above the secondary mirror to further correct the aberrations. The lens is made from fused silica which has a good transmission in the desired band, and is readily available commercially. Primary and secondary baffles will be used to reduce the effect of straylight at the focal plane. The overall obstruction of the telescope is $\lesssim 40\%$.

\section{Mechanical Design}

\begin{figure}[h]
\begin{center}
\includegraphics[scale=0.50]{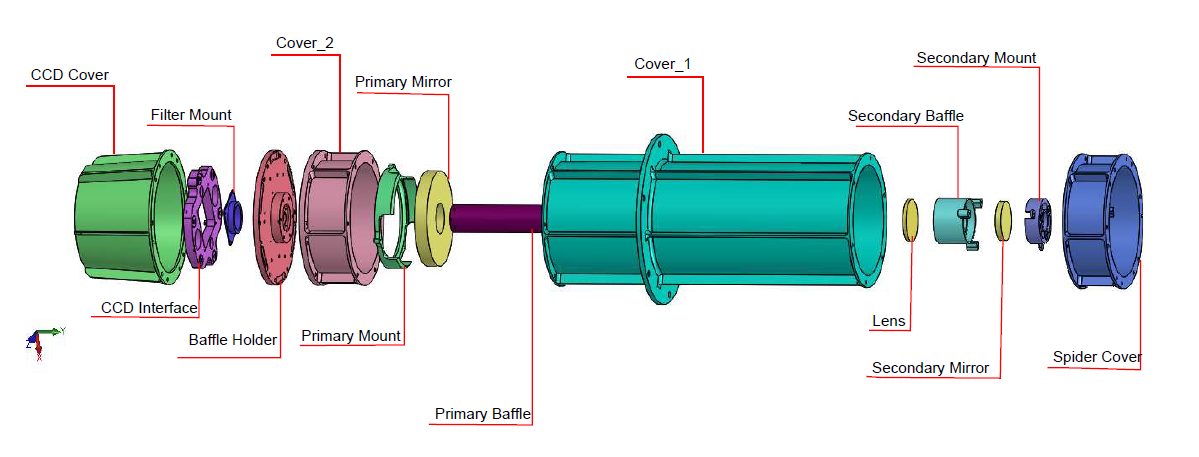} 
\end{center}
\caption{LUCI Structure}
\label{fig:LUCI Structure}
\end{figure}

The instrument is required to be lightweight and compact with a weight constraint of less than 2 kg. The structural design of the instrument is a stiffness-based design to provide a high natural frequency and to withstand the dynamic loads during launch. The instrument has been designed in such a way that it can fly on different platforms, and a generalized simple flange is provided for easy interfacing at the CG (center of gravity) for better dynamic stability. It has been designed with a view of launching it on any future missions and hence higher loads than the other available launch vehicles were considered during the design stage. At earlier stages of the structural design, it was noted that the natural frequency must be above 100 Hz, which is typically the most stringent requirement for most launch vehicle platforms, therefore, a design with high stiffness-to-mass ratio was adapted by using ribs around the main cover. Mechanical requirements are taken to be higher than the launch requirements of the most stringent launcher, so that the instrument can be ready to be flown on any platform. Aluminum alloy 6061-T6 was used as the primary material for the telescope structure due to its high strength-to-density ratio, easy machinability, low cost, availability and low outgassing properties. For the mirror and lens mounts, Invar36 is used due to its very low thermal expansion coefficient and thermal conductivity along with good structural properties as well. The spider for the secondary mirror is a three-vane design, in which the vanes are attached tangentially to the secondary holder. Titanium-6Al-4V, which is a high strength metal alloy, is used for all bolts and washers. The material properties have been assigned according to the Table~\ref{table:Material Properties}. The CAD (computer aided design) model has been made using the mechanical design software SolidWorks \footnote{Solidworks is the mechanical design software. The official website is http://www.solidworks.in/}.The exploded view of the telescope structure is shown in Fig.~\ref{fig:LUCI Structure}.

\begin{table}[h]
\begin{center}
\caption{Properties of the materials used}
\vskip 0.1in
\begin{tabular}{llll} 
%\toprule
Material Property & AA-6061&Invar36&Ti-6Al-4V\\
 \hline
Young's Modulus (MPa) &  68900 &148000&115000 \\
Young's Modulus (MPa) &  68900 &148000&115000 \\
Density (tonne/mm$^3$) & $2.7\times 10^9$ &$8.05\times 10^9$&$4.30\times 10^9$\\
Poisson’s Ratio&  0.33 &0.29& 0.3\\
Tensile Yield Strength&320& 240& 880\\
\hline
\end{tabular}
\label{table:Material Properties}
\end{center}
\end{table}
\section{Detector and Electronics}

Conventional CCD/CMOS detector do not have a sufficient response in the UV and we have chosen the ARTCAM-407 UV camera (Table~\ref{ccd}) in our instrument. This camera uses Sony's ICX407BLA CCD and is specially enhanced for its UV response.

\begin{table}[h]
\caption{Specifications of the UV CCD Camera (Source:ARTRAY CO.,LTD)} 
\label{ccd}
\begin{center} 
\begin{tabular}{l ll} 
 \hline
\rule[-1ex]{0pt}{3.5ex}Sensor &  UV CCD  Array size  6.47(H)x4.83(V) mm \\
\rule[-1ex]{0pt}{3.5ex}Total number of pixels & 1.5 Mp 1360(H)x1024(V)\\
\rule[-1ex]{0pt}{3.5ex}Pixel size & 4.65(H)x4.65(V)μm\\
\rule[-1ex]{0pt}{3.5ex}Frame rate & 12fps\\
\rule[-1ex]{0pt}{3.5ex}Interfac e& Parallel output/USB 2.0\\
\rule[-1ex]{0pt}{3.5ex}Power & 5V\\
\rule[-1ex]{0pt}{3.5ex}Power consumption & 2 W\\
\rule[-1ex]{0pt}{3.5ex}Dimension & 50(W)x47(H)x49.7(D) mm\\

\rule[-1ex]{0pt}{3.5ex}Weight  & 120 gm\\
\hline
\end{tabular}
\end{center}
\end{table} 

\begin{figure}[h]
\begin{center}
\includegraphics[scale=0.9]{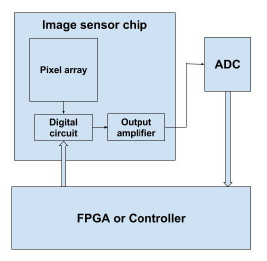} 
\end{center}
\caption{Electronics readout block diagram}
\label{fig:Electronics readout block diagram}
\end{figure}

The basic electronics readout and interface block diagram is shown in  Fig.~\ref{fig:Electronics readout block diagram}.
This camera has a stack of 3 PCBs. First PCB contains the UV CCD and the timing generator chip, which is required to generate reset clocks and transfer clocks required for the CCD. Second PCB contains the voltage regulators to provide various voltages for reset gate, vertical transfer and horizontal transfer clocks. Third PCB contains its own FPGA which transmits the digital data received from the CCD PCB to the computer through USB. This PCB does not store the image data on-board or process it. We have used the CCD PCB and the voltage regulator PCB as is and connected our FPGA board to them. 
The readout data from the CCD will be further processed online by the FPGA and the compressed data will be stored in the on board memory of the system and will be send back whenever data link is available. 

We have developed a generic (application-wise) FPGA board \cite{fpga} to be used as image processor board on any space based image sensor. This board includes a method to generate the clocks to read the image data, and a real-time image processor system which can be used for image processing tasks of different levels. 

\begin{figure}[h]
\begin{center}
\includegraphics[scale=0.9]{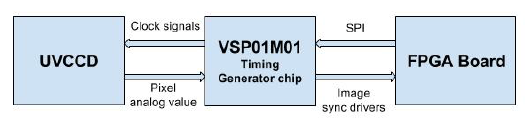} 
\end{center}
\caption{Lay out of UV CCD readout}
\label{fig:lay out of UV CCD readout}
\end{figure}

There is a timing generator chip VSP01M01\footnote{Texas Instruments, USA. {\tt www.ti.com}} along with the image sensor in the same PCB. This chip includes the digital circuitry to access the analog voltage from each pixel and an ADC. It also generates image synchronization driver signals (frame-valid, line-valid, and pixel clock) which are received by the FPGA boardFig.~\ref{fig:lay out of UV CCD readout}. The required timing for various clock signals are to be programmed when circuit is switched on and is done by microbalze in the FPGA. The FPGA board captures the values of each pixels and  decodes the position of the pixel using the synchronization signals from timing generator.

The real time processing in the LUCI which will be carried out by our board , include stray light removing, astrometry, flat fielding, source extraction and cosmic rays rejection, aperture and PSF photometry, data compression and archiving.

\section{Conclusion}

We have described about the design and development a compact UV imaging telescope  which would weigh less than 2 kg to study the bright UV sources. The instrument can fly on a different platforms and our first experiment would be on a balloon platform which we are planning later this year. The main constraints in the design of the complete payload are its compactness, light weight and cost effectiveness, and we were able to meet all these challenges.

\acknowledgements
Part of this research has been supported by the Department of Science and Technology (Government of India) under Grant IR/S2/PU-006/2012.
 
%\vskip lin

\end{document}